\documentclass[12pt,letterpaper]{article}                  
\usepackage{osajnl}
\usepackage{overcite,hyperref}

\begin{document}

\title{Controlled switching of discrete solitons in waveguide arrays}

\author{Rodrigo A. Vicencio and Mario I.  Molina}

\affiliation{Departamento de F\'{\i}sica, Facultad de Ciencias,
Universidad de Chile, Casilla 653, Santiago, Chile}

\author{Yuri S. Kivshar}

\affiliation{Nonlinear Physics Group, Research School of Physical
Sciences and Engineering, The Australian National University,
Canberra ACT 0200, Australia}

\vspace{2cm}

\begin{abstract}
We suggest an effective method for controlling nonlinear switching
in arrays of weakly coupled optical waveguides. We demonstrate the
digitized switching of a narrow input beam for up to eleven
waveguides in the engineered waveguide arrays.
\end{abstract}


\maketitle
\newpage

Discrete optical solitons were first suggested theoretically as
stationary nonlinear localized modes of a periodic array of weakly
coupled optical waveguides~\cite{chrjos88}. Because the use of
discrete solitons promises an efficient way to control multi-port
nonlinear switching in a system of many coupled waveguides, this
field has been extensively explored
theoretically~\cite{kiv93,kro_kiv96,ace_et96,ieee_review}. More
importantly, the discrete solitons have also been generated
experimentally in fabricated periodic waveguide
structures~\cite{eis_et98,silste01}.

The most common theoretical approach to study discrete optical
solitons in waveguide arrays is based on the decomposition of the
electric field in the periodic structure into a sum of weakly
coupled fundamental modes excited in each waveguide of the array.
According to this approach, the wave dynamics can be described by
an effective discrete nonlinear Schr\"odinger (DNLS) equation,
that possesses spatially localized stationary solutions in the
form of localized modes of a lattice model. Many properties of the
discrete optical solitons can be analyzed in the framework of the
DNLS equation~\cite{chrjos88,kiv93,kro_kiv96,ace_et96}.

One of the major problems for achieving the controllable
multi-port steering of discrete optical solitons in waveguide
arrays is the existence of an effective periodic potential which
appears due to the lattice discreteness, known as the
Peierls-Nabarro (PN) potential. It represents the energy cost
associated with a shift of a nonlinear localized mode by a half of
the waveguide spacing~\cite{morand_99}. Its magnitude can be roughly
estimated as $\sim |A|^{4}$, where $A$ is the soliton amplitude. 
As a consequence of this
potential, a narrow large-amplitude discrete soliton does not
propagate in the lattice and it becomes trapped by the array.
Several ideas to exploit the discreteness properties of the array
for the optical switching were suggested \cite{aceves94,bang96},
including the demonstration of the output channel selection for
the multi-port devices. However, the soliton steering and
switching is well controlled only in the limit of broad beams
whereas the soliton dynamics in highly discrete arrays has been
shown to be more complicated \cite{bang96}. In this Letter, we
suggest a ``discreteness engineering'' approach and demonstrate how
to achieve highly controllable multi-port soliton switching in the
arrays by a desired integer number of waveguides, the so-called
``digital soliton switching''.

We consider a standard model of the waveguide arrays with a
modulated coupling described by the normalized DNLS equation of
the form,
\begin{equation}
\label{eq_1}
 i\frac{du_n}{dz} + V_{n+1} u_{n+1} + V_{n-1} u_{n-1}
+ \gamma |u_n|^2 u_n = 0,
\end{equation}
where $u_n$ is the effective envelope of the electric field in the
$n$-th waveguide and $z$ is the propagation distance. Unlike the
standard models~\cite{chrjos88,kiv93,kro_kiv96,ace_et96}, the
coupling $V_n$ between two neighboring guides is assumed to vary,
either through the effective propagation constant or by changing
the spacing between neighboring guides. The parameter $\gamma =
\omega_0 n_2/(c A_{\rm eff})$ is the effective waveguide
nonlinearity associated with the Kerr nonlinearity of the core
material.

The steering and trapping of discrete solitons in the framework of
the model (\ref{eq_1}) have been analyzed in many studies. Being
kicked by an external force, the discrete soliton can propagate
through the lattice for some distance, but then it gets trapped
due to the effect of discreteness. For a larger force, the output
soliton position fluctuates between two (or more) neighboring
waveguides making the switching uncontrollable~\cite{bang96}.
Here, we suggest to modulate the waveguide coupling in order to
achieve a controllable output and to engineer the switching
results. {\em The key idea is to break a symmetry between the beam
motion to the right and left at the moment of trapping}; this
allows the elimination of chaotic trapping observed in homogeneous
arrays~\cite{bang96}. In this way, we achieve a controllable
digitized switching where the continuous change of the input beam
amplitude results in a quantized displacement of the output beam
by an integer number of waveguides.

We have tested different types of modulation in the array
parameters and the corresponding structures of array
super-lattices. An example of one of the optimized structures is
shown in Fig. 1, where we modulate the coupling parameter
$V_n$ in a step-like manner. We also notice that the use of a
linear ramp potential (e.g. of the form $V_n = a n$) for this 
purpose does not lead to an effective switching but, instead,
makes the soliton switching even more chaotic due to the
phenomenon of Bloch oscillations which become randomized in the
nonlinear regime~\cite{morandotti2_99}.

We select the input profile in the form
of a narrow sech-like beam localized only on a few waveguides,
\begin{equation}
\label{eq_in} u_n(0) = A\ {\rm sech} [A(n-n_c)/\sqrt{2}]
\ e^{-ik(n-n_c)},
\end{equation}
for $n-n_{c}=0, \pm1$, and $u_n=0$, otherwise. For the particular
results presented below, we select the array of $101$ waveguides and
place the beam at the middle position, $n_c =50$. The maximum
normalized propagation distance used in our simulations is $z_{\rm
max} = 45$ (in units of the coupling length).

Parameter $k$ in the ansatz (\ref{eq_in}) has the meaning of the
transverse steering velocity of the beam, in analogy with the
continuous approximation. It describes the value of an effective
kick of the beam in the transverse direction at the input, in
order to achieve the beam motion and shift into one of the
neighboring (or other desired) waveguide outputs.

In our simulations, we control the numerical accuracy by
monitoring two conserved quantities of the model (\ref{eq_in}),
the soliton power $P = \sum_{n} |u_n(z)|^2$, and the system
Hamiltonian, $H = \sum_n \left\{ V_n (u_n u_{n+1}^* + u_n^*
u_{n+1}) + (\gamma/2) |u_n|^4 \right\}$.

The input condition (\ref{eq_in}) is not an exact stationary
solution of the discrete equation (\ref{eq_1}) even for $k=0$, and
as the input kick ($k \neq 0$) forces the soliton to move to the
right ($k<0$) or left ($k>0$), the motion is accompanied by some
radiation. The effective lattice discreteness can be attributed to
an effective periodic PN potential. Due to both the strong
radiation emission and the PN barrier which should be overtaken in
order to move the beam, the discrete soliton gets trapped at one
of the waveguides in the array, as shown in Fig. 2. In most cases,
the shift of the beam position to the neighboring waveguide is
easy to achieve, as shown in many studies \cite{bang96}. However,
the soliton switching becomes rather complicated and even chaotic
when the kicking force becomes stronger.

We have studied many different regimes of the soliton multi-port
switching in the array and revealed that the most effective
switching in a desired waveguide position (i.e. desired output)
can be achieved by varying the coupling between waveguides, either
through the effective propagation constant or by changing the
spacing between neighboring guides, as shown in Fig. 1. This
$V_{n}$ profile was obtained after performing a numerical sweep in
$V_{n}$ and $A$ for fixed momentum $k$. In this case, the
selection  of a finite value of the steering parameter $k$ allows
to switch the whole beam into a neighboring waveguide, as shown in
Fig. 2, with only a small amount of radiation. By decreasing the
amplitude of the input pulse at a fixed value of the steering
parameter, fixed to be say $k =\pm 0.9$, it is possible to achieve
self-trapping of the soliton beam at some (short) distance from
the initial center at different waveguide position. Due to the
step-like modulated coupling, we create a selection between the
beam motion to the right and left at the moment of trapping thus
suppressing or eliminating the chaotic trapping observed in
homogeneous waveguide arrays. In this way, we achieve a
controllable digitized nonlinear switching where the continuous
change of the amplitude of the input beam results in a quantized
displacement of the output beam by an integer number of
waveguides. Consequently, for the parameters discussed above we
observe almost undistorted switching up to eleven waveguides, and
Fig. 3 shows an example of the digital soliton switching to the
eleventh waveguide \cite{comment}.

Figure 4 gives a summary of the results for the parameters
discussed above; it shows the discrete position of the soliton at
the output as a function of the input beam amplitude, for two
fixed values of the steering parameter $k = \pm 0.9$. In a remarkable
contrast with other studies, the coupling modulation allows a
controllable digitized switching in the array with very
little or no distortion. The figure also shows a slight asymmetry
in the final displacement, depending on whether the beam is kicked
uphill or downhill.

If we were to use five guides instead of three for the input beam,
one could expect a smaller amount of radiation emitted. However,
this would imply a longer distance before the beam gets trapped by
one of the waveguides in the array due to the effective
Peierls-Nabarro potential. Also, this means that one could in
principle switch the soliton beam to any desired waveguide in the
waveguide array, no matter how far; it would be just a matter of
choosing an initial beam wide enough, i.e., closer to the
continuum limit (in addition to the optimization of the coupling
in a step-wise manner) by removing the random selection between
the directions and suppressing the beam random switching.

Another observation is that the sech-like initial profile is not
really fundamental.  A (kicked) nonlinear impurity-like profile of
the form $u_{n}(0) = A [ (1-A^{2})/(1 + A^2) ]^{|n-n_{c}|/2}\
\exp[-i k (n-n_{c})]$ will also show similar behavior, as our
additional computations show. The reason for this behavior seems
to rest on the observation that, for any system with local
nonlinearity, a narrow initial profile will effectively render
the system into a linear one containing a small nonlinear cluster
(or even a single site); the bound state will therefore strongly
resemble the one corresponding to a nonlinear impurity\cite{last}.

In conclusion, we have suggested a novel approach to achieve a
digitized switching in waveguide arrays by using the concept of
discrete optical solitons. Our approach involves a weak step-like
modulation of the coupling strength (or, equivalently, distance
between the waveguides) in the arrays with the period larger than
the waveguide spacing. Such a super-lattice structure allows
the modification of trapping properties of the waveguide array due to
discreteness which in turn permits engineering of the strength of the
effective trapping
potential. We have
demonstrated numerically the controlled switching up to
eleven waveguides in the arrays by using very narrow input beams
localized on three waveguides only.

R. A. Vicencio acknowledges support from a Conicyt doctoral
fellowship. M.I. Molina and Yu. S. Kivshar acknowledge support
from Fondecyt grants 1020139 and 7020139. In addition, Yu.S.
Kivshar acknowledges the warm hospitality of the Department of
Physics of the University of Chile.

\newpage

\centerline{\bf Figure Captions}
\vspace{2cm}

Fig. 1. Example of the optimized modulation of the propagation
constant $V_n$ in the waveguide array. \vspace{0.5cm}

Fig. 2. One-site switching of a discrete soliton in the waveguide
array with the modulated coupling shown in Fig. 1. \vspace{0.5cm}

Fig. 3. Discrete switching by eleven sites
 in the waveguide array modulated according to Fig. 1.
\vspace{0.5cm}

Fig. 4. Soliton switching in a waveguide array with an optimized
coupling. Shown is the soliton output displacement as a function
of the input beam amplitude (A). A step size of 0.0005 separates
consecutive points.

\newpage

\begin{figure}[h]\centerline{\scalebox{1}{\includegraphics{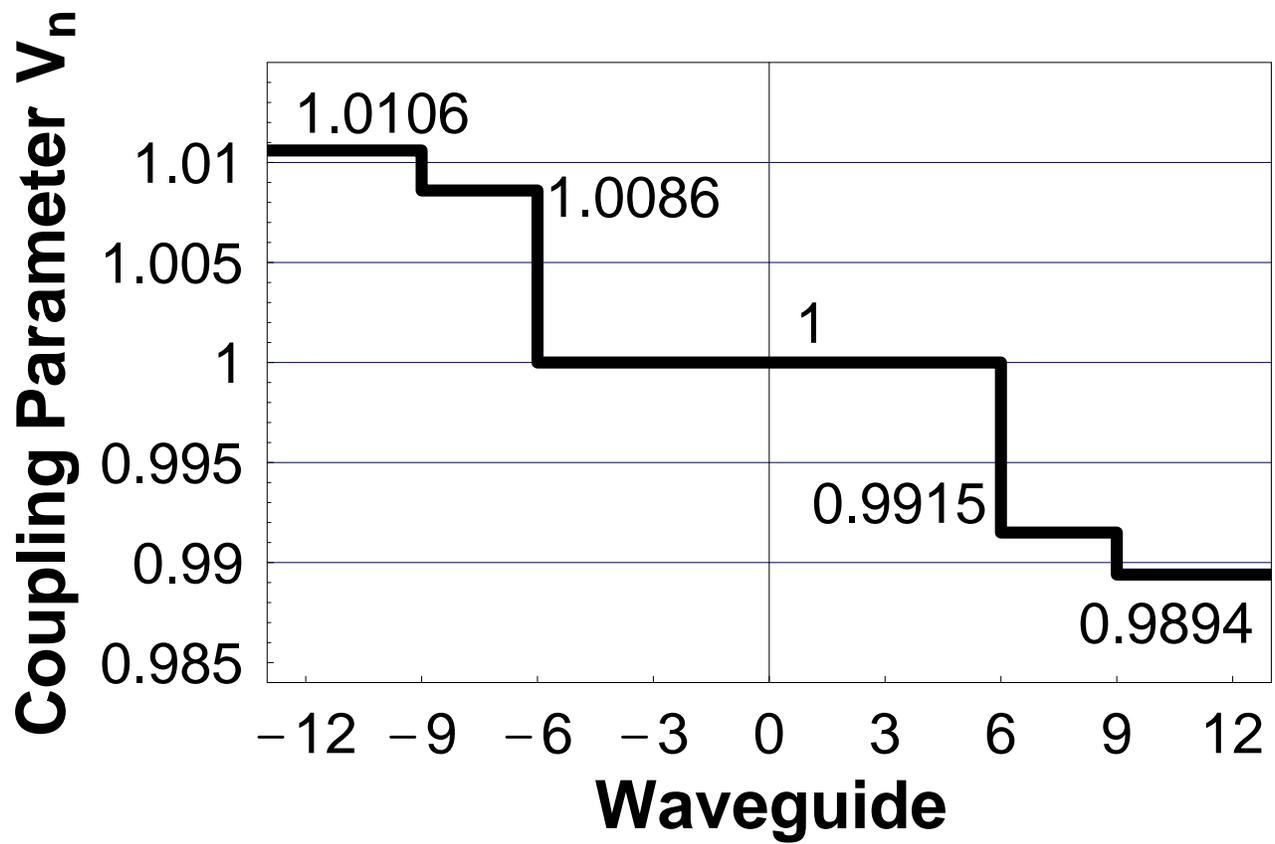}}}
\label{fig1} \caption{Example of the optimized modulation of the
propagation constant $V_n$ in the waveguide array.}
\end{figure}

\begin{figure}[h]\centerline{\scalebox{1}{\includegraphics{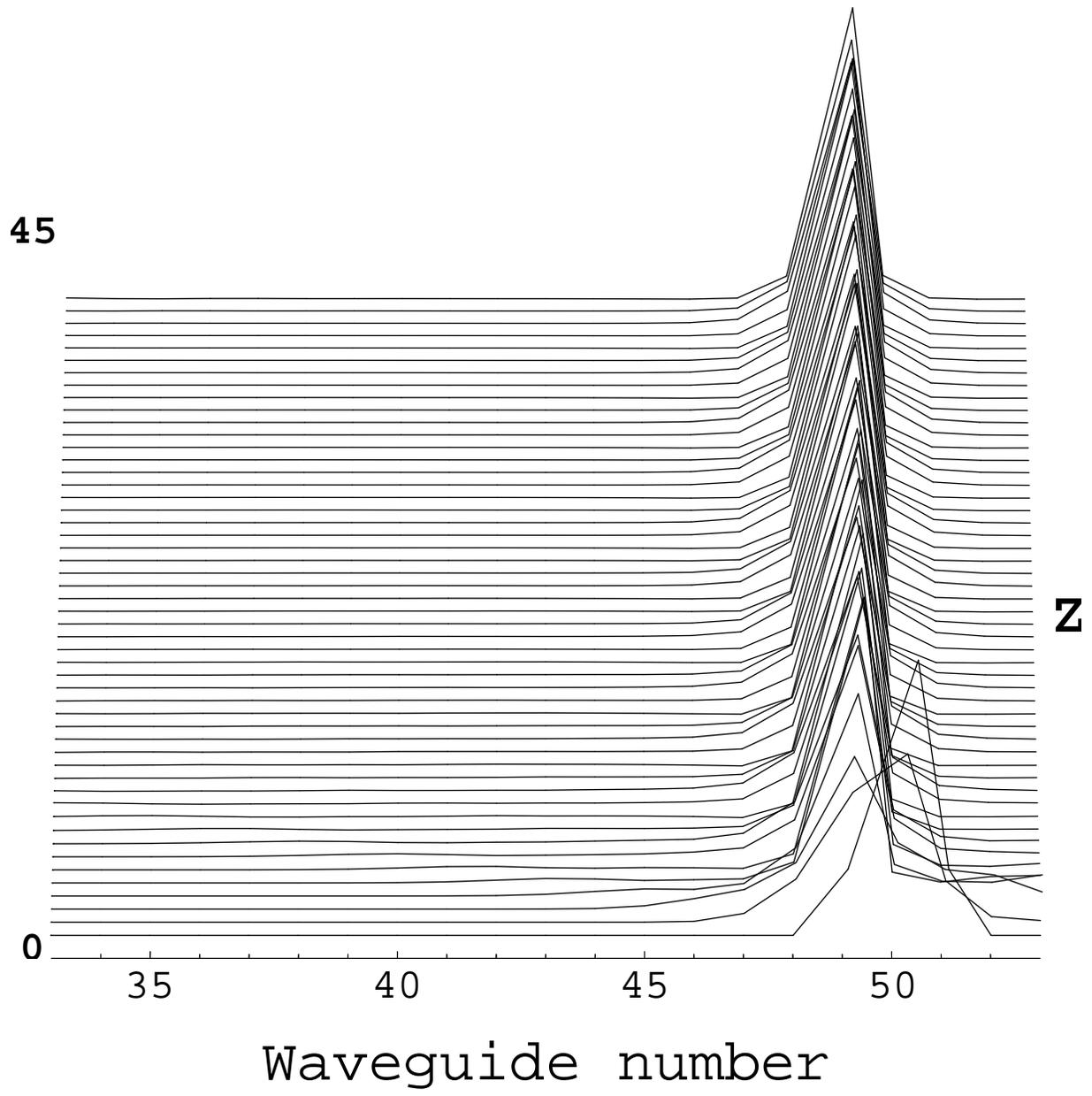}}}
\label{fig2} \caption{One-site switching of a discrete soliton in
the waveguide array with the modulated coupling shown in Fig. 1.}
\end{figure}

\begin{figure}[h]\centerline{\scalebox{1}{\includegraphics{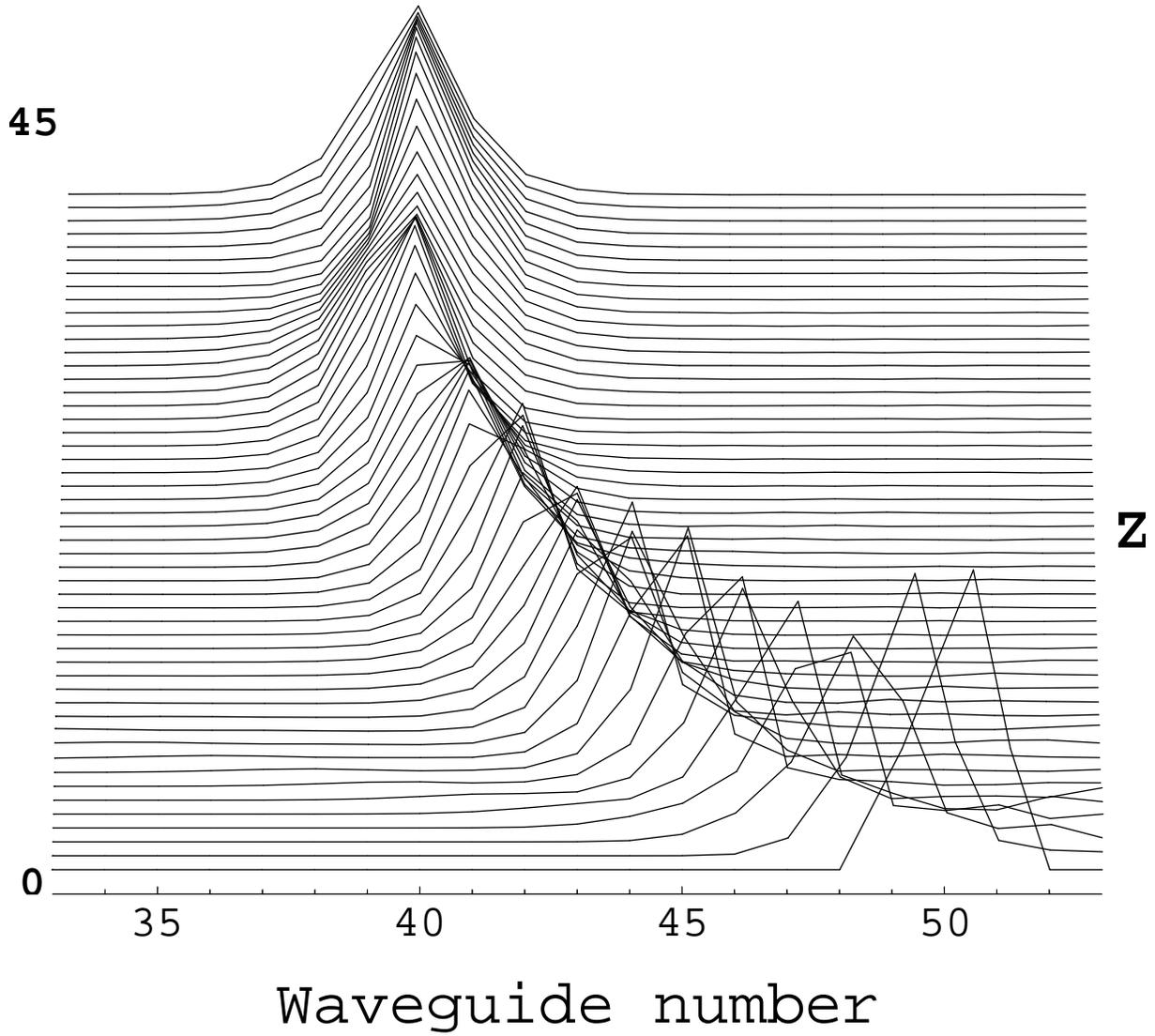}}}
\label{fig3}
\caption{Discrete switching by eleven sites
in the waveguide array modulated according to Fig. 1.}
\end{figure}

\begin{figure}[h]\centerline{\scalebox{1}{\includegraphics{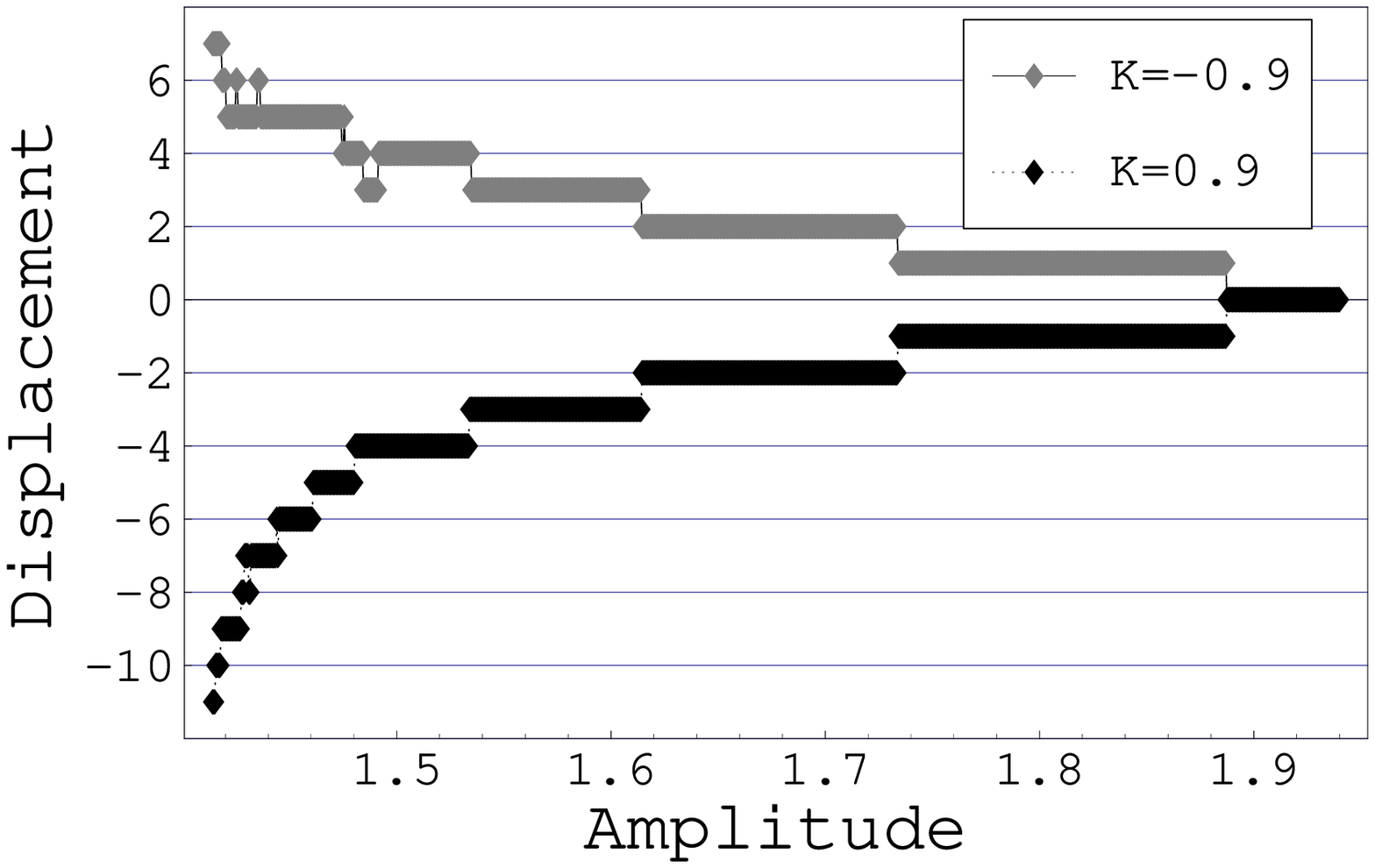}}}
\label{fig4}
\caption{Soliton switching in a waveguide array with an optimized
coupling. Shown is the soliton output displacement as a function
of the input beam amplitude $A$.}
\end{figure}


\begin{thebibliography}{}
\bibitem{comment}
The use of a symmetric $V_n$ around $n = n_{c}$ allows to reach
eleven guides on either side, thus making possible switching up to
twenty-two guides.
\end{thebibliography}
\end{document}